\documentclass[runningheads]{llncs}
\usepackage{graphicx}
\usepackage{verbatim} 

\newcommand\figref[1]{Figure~\ref{fig:#1}}
\newcommand\secref[1]{Section~\ref{sec:#1}}

\newcommand\tabref[1]{Table~\ref{tab:#1}}

\usepackage{xcolor}

\newcommand\Nm[1]{Nm$^{-1}$} 
\newcommand\ms[1]{ms$^{-1}$} 
\newcommand\mss[1]{ms$^{-2}$}

\newcommand\sigaccx[1]{$\sigma_\mathrm{Acc, x}$} 
\newcommand\sigaccy[1]{$\sigma_\mathrm{Acc, y}$}
\newcommand\kuraccx[1]{$\kappa_\mathrm{Acc, x}$} 
\newcommand\kuraccy[1]{$\kappa_\mathrm{Acc, y}$}
\newcommand\skewaccx[1]{$\gamma_\mathrm{Acc, x}$} 
\newcommand\skewaccy[1]{$\gamma_\mathrm{Acc, y}$} 
\newcommand\sigposx[1]{$\sigma_\mathrm{Dis, x}$} 
\newcommand\sigposy[1]{$\sigma_\mathrm{Dis, y}$}
\newcommand\sigpostot[1]{$\sigma_\mathrm{tot}$}

\newcommand\kurposx[1]{$\kappa_\mathrm{Dis, x}$} 
\newcommand\kurposy[1]{$\kappa_\mathrm{Dis, y}$}
\newcommand\skewposx[1]{$\gamma_\mathrm{Dis, x}$} 
\newcommand\skewposy[1]{$\gamma_\mathrm{Dis, y}$} 

\newcommand\utot[1]{$u_\mathrm{tot}$}
\newcommand\ux[1]{$u_\mathrm{x}$}
\newcommand\uy[1]{$u_\mathrm{y}$}

\newcommand\fposx[1]{$f_\mathrm{Dis, x}$}
\newcommand\fposy[1]{$f_\mathrm{Dis, y}$}
\newcommand\dist[1]{$d$}
\newcommand\degree[1]{$^{\circ}$}

\begin{document}
\title{Data-driven prediction of vortex-induced vibration response of marine risers subjected to three-dimensional current} 
\titlerunning{Data-driven prediction of VIV in 3D}
%
\author{Signe Riemer-S\o{}rensen\inst{1}\orcidID{0000-0002-5308-7651} \and
Jie Wu\inst{2} \and
Halvor Lie\inst{2} \and
Svein S\ae{}vik \inst{3}\orcidID{0000-0001-5950-6186} \and
Sang-Woo Kim\inst{3}
}
\authorrunning{S. Riemer-S\o{}rensen et al.}
%
\institute{Department of Mathematics and Cybernetics, SINTEF Digital, Oslo, Norway \email{signe.riemer-sorensen@sintef.no} \\ \url{} 
\and {Department of Engergy and Transport, SINTEF Ocean, Trondheim, Norway} \email{jie.wu@sintef.no, halvor.lie@sintef.no}\\
\and Department of Marine Technology, NTNU, Trondheim, Norway \email{svein.savik@ntnu.no, sangwoo.kim@ntnu.no} \\
}
\maketitle              
%
\begin{abstract}
Slender marine structures such as deep-water marine risers are subjected to currents and will normally experience Vortex Induced Vibrations (VIV), which can cause fast accumulation of fatigue damage. The ocean current is often three-dimensional (3D), i.e., the direction and magnitude of the current vary throughout the water column. 

Today, semi-empirical tools are used by the industry to predict VIV induced fatigue on risers. The load model and hydrodynamic parameters in present VIV prediction tools are developed based on two-dimensional (2D) flow conditions, as it is challenging to consider the effect of 3D flow along the risers. Accordingly, the current profiles must be purposely made 2D during the design process, which leads to significant uncertainty in the prediction results. 

Further, due to the limitations in the laboratory, VIV model tests are mostly carried out under 2D flow conditions and thus little experimental data exist to document VIV response of riser subjected to varying directions of the current. However, a few experiments have been conducted with 3D current. We have used results from one of these experiments to investigate how well 1) traditional and 2) an alternative method based on a data driven prediction can describe VIV in 3D currents. 

Data driven modelling is particularly suited for complicated problems with many parameters and non-linear relationships. We have applied a data clustering algorithm to the experimental 3D flow data in order to identify measurable parameters that can influence responses. The riser responses are grouped based on their statistical characteristics, which relate to the direction of the flow. Furthermore we fit a random forest regression model to the measured VIV response and compare its performance with the predictions of existing VIV prediction tools (VIVANA-FD).

\keywords{Machine learning  \and Marine risers \and Vortex induced vibrations.}
\end{abstract}
\section{Introduction}
\begin{figure}[bt]
\centering
\includegraphics[width=0.70\textwidth]{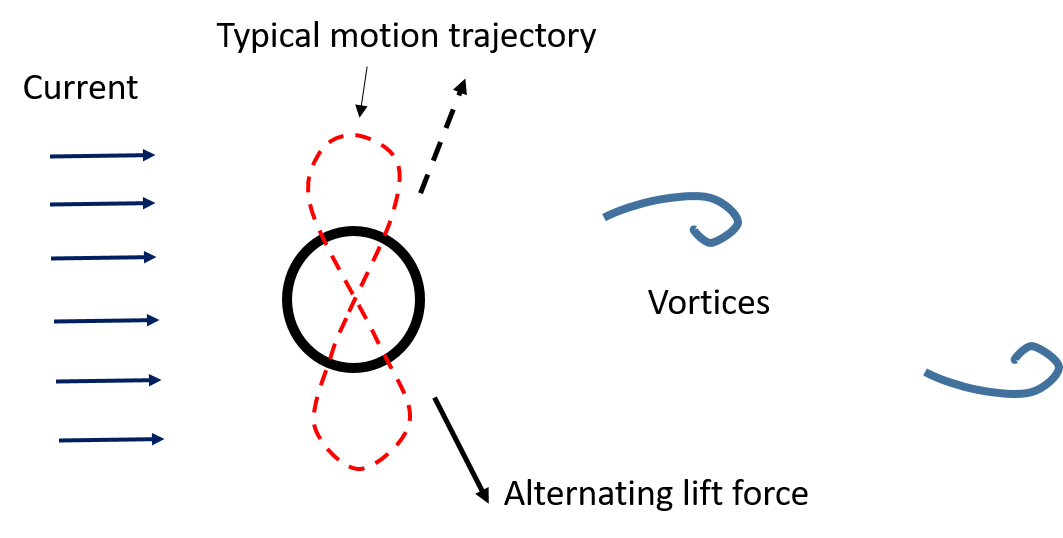}
\hspace{4mm}
\unskip\ \vrule\ 
\includegraphics[width=0.15\textwidth]{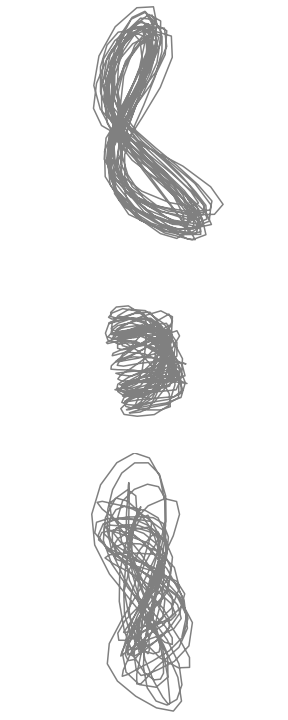}
\caption{{\it Left panel:} Vortex Induced Vibrations (VIV) due to vortices shed in the wake of a slender marine riser (seen from above). The VIV manifest as movements e.g. in a figure-of-eight like pattern (dashed red line). {\it Right panel:} Examples of riser trajectories in the x/y plane perpendicular to the length of the riser.} \label{fig:viv}
\end{figure}

Slender marine structures such as deep water marine risers are exposed to ocean currents causing vortices to be shed in the wake of the circular cross-section as illustrated in \figref{viv}. This will cause alternating lift forces that may synchronise with the cylinder's motion such that high-frequency vibrations can occur, a phenomenon termed Vortex Induced Vibrations (VIV). The vortex shedding frequency may synchronise to a multiple set of riser eigen-frequencies. This may lead to fatigue during short exposure times even if the associated response amplitudes are small. As this limits riser lifetime, VIV are a major concern in riser design and operation.

The load model and hydrodynamic parameters in present frequency domain VIV prediction tools used by industry, i.e., VIVANA \cite{Passano2014}, Shear7 \cite{Vandiver2007} and VIVA \cite{Triantafyllou1999}, are developed for the simplified 2D flow conditions. However, the speed of the current and spatial patterns in the field may vary over the water column leading to 3D flow conditions \cite{Srivilairit2009}. It is common practice to perform pure cross-flow VIV analysis with current profiles purposely made 2D, which is an obvious over-simplification. This is partially due to the lack of a reliable model in present VIV prediction algorithms of combined in-line and cross-flow responses. 
In addition, the hydrodynamic load in the frequency domain prediction tools is assumed to be harmonic, which is a simplification of the true load process (non-harmonic). A new time domain prediction tool has been developed \cite{Thorsen2014,Thorsen2017}, but systematic validation of the tool subjected to 3D flow condition is needed.

Experimental VIV model tests subjected to 2D flow have been carried out in recent years \cite[e.g.]{Lie2012,TRIM2005335,Vandiver2006}. Limited test data can be found for VIV in 3D flow conditions. The common challenge of these data is that VIV response becomes increasingly complex with many parameters that interplay and influence the physical process. Consequently, it is also difficult to model the process in a simplified mathematical model.

Data driven modelling and machine learning algorithms are powerful and highly flexible model-free methods for inferring relationships in data \cite{Hastie09}. They are constructed to handle highly complex problems with many parameters and non-linear relationships. However, to take full advantage of the methods, high-quality data with many samples are required.

VIV response modelling is complex and model tests provide experimental data for testing the suitability of a machine learning approach (\secref{data}). First we applied unsupervised learning (clustering) to explore and identify relationships in the data in order to select the most relevant parameters from the high-dimensional data set (\secref{clustering}). Secondly, we used supervised learning to build a regression model that can predict the statistical properties of the VIV response based on the flow properties (\secref{supervised}).

The overall objective of the research is to obtain more accurate estimation of riser VIV fatigue damage by combining traditional prediction methods with machine learning to be integrated in future on-site riser monitoring systems. This paper presents a first attempt on the possible improvements by including the effect of 3D current in the otherwise 2D traditional VIV prediction tools. VIV response prediction is carried out with existing VIV prediction tools and compared with similar predictions from a data driven model in order to investigate the limitations and potentials of both methods. 


\section{VIV model test with 3D flow} \label{sec:data}
We use data from an experiment carried out in 1996 and 1997 at the MARINTEK (now part of SINTEF Ocean) towing tank using a rotating rig. 

\subsubsection{Experimental setup and riser model:}
\begin{figure}[bt]
\centering
\includegraphics[width=0.9\textwidth, trim={0 0 0 0}, clip]{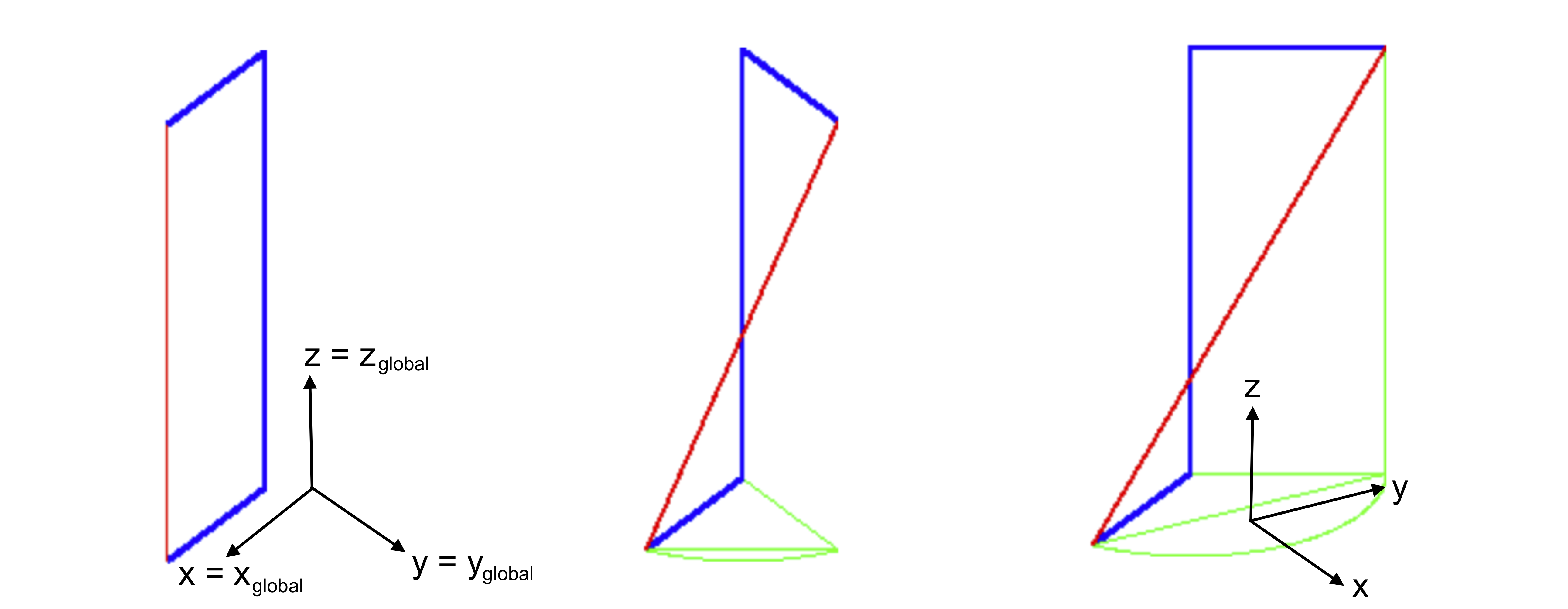}
\includegraphics[width=0.9\textwidth]{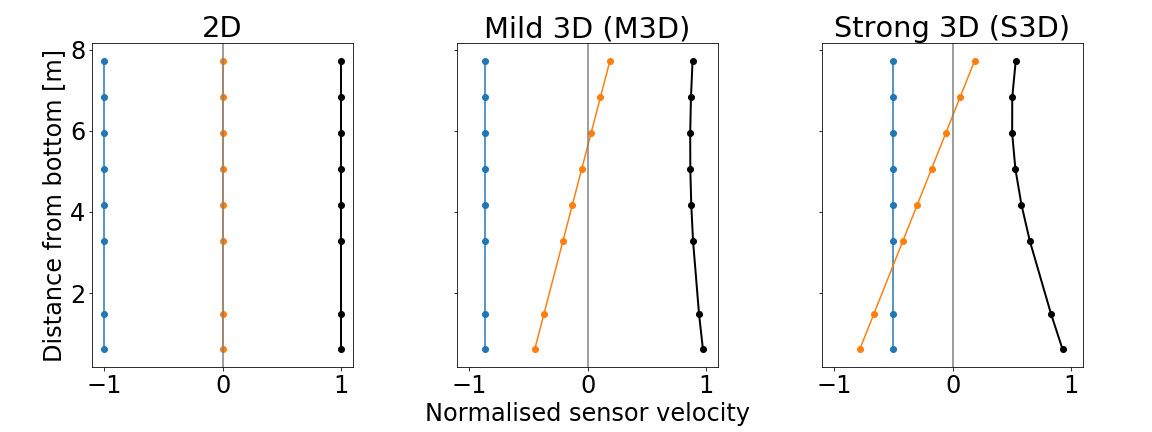}
\caption{{\it Upper panel:} The geometrical configuration of the arms (blue) and the riser (red). In the 2D configuration (left), the angle between the horizontal arms was 0\degree{} leading to a 2D flow. In the mild 3D configuration (middle) the angle between the arms was 60\degree{} and in the strong 3D configuration (left) the angle was 120\degree{}, both leading to a 3D sheared flow. {\it Lower panel:} Normalised velocities along the riser in the x-direction ($u_x/U$, blue), y-direction ($u_y/U$, orange), and magnitude of total velocity (\utot{}$/U$, black thick line), where $U$ refers to the velocity of the tip of the lower arm.} \label{fig:geometry}
\end{figure}

In the laboratory, 3D flow conditions were mimicked using a rotation rig as shown in \figref{geometry}. The rig was mounted in the deepest part of a towing tank having a length, width, and depth of 80\,m, 10.5\,m and 10\,m, respectively. The test rig consisted of a 13\,m long vertical cylinder with a diameter of about 0.5\,m. Bearings in both ends of the cylinder made it possible to rotate the rig around its vertical axis by use of an electrical actuator. The rotating cylinder was mounted with horizontal arms at the top and bottom. The riser model was suspended between these arms in a pretension arrangement. At the lower connection point it had a constant arm length of 4.6\,m. The upper connection arm length could be varied from 0.4\,m to 4.6\,m. 

During the test, the rig was rotated around the vertical cylinder at constant rotational velocity. Thus, the riser experienced a water flow given by the rotational speed and distance to the rotational axis. For equal arm lengths, the flow speed was uniform in magnitude and direction along the riser model. Decreasing the upper arm length, led to a variation in the flow along the riser model, resulting in a sheared, but still 2D, flow. More detailed description of the test set-up can be found in \cite{Lie1998}.

In order to accommodate 3D flow conditions, the planar angle between the upper and lower arms could be adjusted.
When the two arms were offset, the current flow varied in direction along the riser model. Hence, it was possible to test a riser model subjected to a well-defined 3D current profile. Seven different setup set-up configurations were performed, where the length of the upper arm varied between 2.7\,m to 4.6\,m, and the angle between the horizontal arms varied between 0\degree{} and 165\degree{}. In the present study only the three configurations with constant upper arm length of 4.6\,m were included. The angle between the arms were 0\degree{}, 60\degree{} and 120\degree{}, resulting in a uniform 2D current profile, a mild 3D current profile and a strong 3D current profile. 

The outer diameter of the test model riser was 23\,mm with a weight in water of 1.433\,\Nm. The riser had been modified to house accelerometers, and stiffened with a 4\,mm steel wire to obtain a high axial elastic stiffness and strength. It was filled with gelatin to avoid vibrations of the cabling inside.

\subsubsection{Time series:} \label{sec:timeseries}
The response of the riser was measured with 10 pairs of bi-axial accelerometers mounted along the length of the riser. The accelerations were measured along and perpendicular to the length of the riser with a sampling frequency of 120\,Hz. During the tests two of the accelerometers (number 3 and 10 from the bottom) failed and were rejected from further analysis.

\begin{figure}[bt]
\centering
\includegraphics[width=\textwidth]{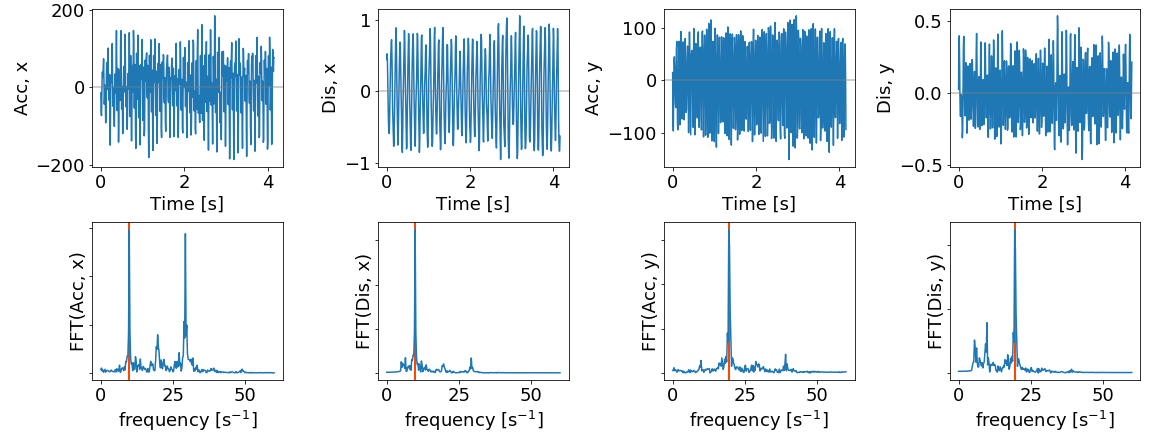}
\caption{An example of a time series interval for sensor 5 in the 2D geometry with rotational velocity of 1.629\,\ms{} in the sample interval 5000 to 5500. The upper panels show the time series and the lower panels the real part of the fast Fourier transform.}
 \label{fig:timeseries_fft}
\end{figure}

The measured accelerations were Fourier transformed and low-pass filtered at 30\,Hz to remove high frequency noise, before being doubly integrated in the frequency domain to obtain displacement time series. The transient phase in the measured signals at the beginning and the end of each test case were discarded so the total signal consisted of time steps from 2500 to 8000 (21.8--66.7\,s). We split the time series in intervals of 500 samples corresponding to 4.2\,s an example of which is given in \figref{timeseries_fft}. We used a fast Fourier transform of the time intervals to get the dominating oscillation frequencies (lower panels of \figref{timeseries_fft}). 

We only considered local current and treated each sensor and each time interval as independent measurements. This is a strong simplification since it is known that VIV responses will be correlated over the length of the riser model.

The inaccuracy of the accelerometers and other sensors was a few percent of the true values. In addition, it is observed that VIV response is not completely stationary. Therefore, the standard deviation of the measured response can vary depending on the time slot. For the total displacement, the deviation between the time slots can be up to 15\%.

\subsubsection{Features:}
In total we considered the features listed in \tabref{params}. They are not independent and at different stages in the analysis different subsets were selected as specified.

In addition to the variance (standard deviation) of the distributions, we also regarded skewness and kurtosis. Skewness is the third standardised central moment for the probability distribution. It describes the symmetry of the distribution. Kurtosis is the fourth standardised central moment of the probability distribution. It is a measure of the combined weights of the tails relative to the rest of the distribution \cite{Westfall2014}. Vandiver \cite{Vandiver2000} used kurtosis to characterise the VIV responses. It assumes a value of 1.5 for a sinusoidal process as typical for a single mode lock-in response, and a value of 3.0 for a Gaussian process typical for multi-frequency random vibration. 

\begin{table}[tb]
\caption{The full set of considered features. The x- and y- directions refer to the plane perpendicular to the length of the riser (see \figref{geometry})} \label{tab:params}
\begin{tabular}{l | l}
\hline
Feature & Explanation\\
\hline
\sigaccx{}, \sigaccy{}			& Standard deviation of acceleration in the x, y directions in \mss{}\\
\skewaccx{}, \skewaccy{}		& Skewness of acceleration in the x, y direction, no units  \\
\kuraccx{} , \kuraccy{} 		& Kurtosis of acceleration in the x, y direction, no units  \\
\sigposx{}, \sigposy{} 		    & Standard deviation of displacement divided by riser diameter \\
\sigpostot{}                    & Standard deviation of total displacement divided by riser diameter \\
\skewposx{}, \skewposy{}		& Skewness of displacement in the x, y direction, no units  \\
\kurposx{} , \kurposy{} 		& Kurtosis of displacement in the x, y direction, no units  \\
\fposx{}, \fposy{} 			& Frequency of oscillations in x, y direction in units of s$^{-1}$  \\
\utot{}					& Total sensor velocity (equal to current) in \ms{} \\
\ux{}, \uy{}					& Magnitude of sensor velocity in the x, y direction in \ms{}\\
$U$                             & Velocity of lower arm in \ms{}\\
\dist{}					& Sensor distance from bottom in m \\
\hline
\end{tabular}
\end{table}

\section{Clustering} \label{sec:clustering}
We used unsupervised learning to investigate the relations between the dimensionality of the current and the statistical properties of the oscillations. In order to focus on the flow dimensionality, we only considered measurements where the total sensor velocity fell within a narrow range, \utot{}$=1.1-1.2$\,\ms{}.

\subsubsection{Hierarchical density based clustering:}
Clustering basically means sorting the data points according to similarities in the full parameter space. For a few features this can easily be visualised, but for many features visualisation is challenging.

Hierarchical density based clustering uses the density of the points in feature space to build cluster trees hierarchically. The tree is then condensed based on a minimum cluster size \cite{Campello2013,McInnes2017}. Contrary to other clustering methods, the user does not specify a number of clusters, but instead specifies the minimal cluster size. The points that do not fulfil the criteria for becoming a cluster are assigned to a noise/outlier category. The user can control the conservatism of the clustering i.e. how strict the algorithm should be when assigning points to clusters or noise.
Density based scanning is particularly good at handing elongated and overlapping clusters, and due to the noise option it is suitable for exploratory data analysis.

\subsubsection{Method and results of clustering:}
We applied hierarchical density based clustering to the full set of features in \tabref{params} with Euclidian distance as similarity matrix, minimum cluster size of 18 members, and minimum number of samples (the conservatism) of 1. Since the data were equally distributed among the three scenarios with different governing physics, we expected the clusters to be small and homogeneous, and consequently we used the leaf method for cluster selection (rather than the default mass excess method which has a tendency to produce large clusters).

The importance of individual features in the clustering combined with physical reasoning and correlations between the features were used to select a subset of relevant features consisting of: \sigaccx{}, \sigaccy{}, \kuraccx{}, \kuraccy{}, \fposx{}, \fposy{}, \ux{}, \uy{}.

\begin{figure}
\centering
\includegraphics[width=\textwidth]{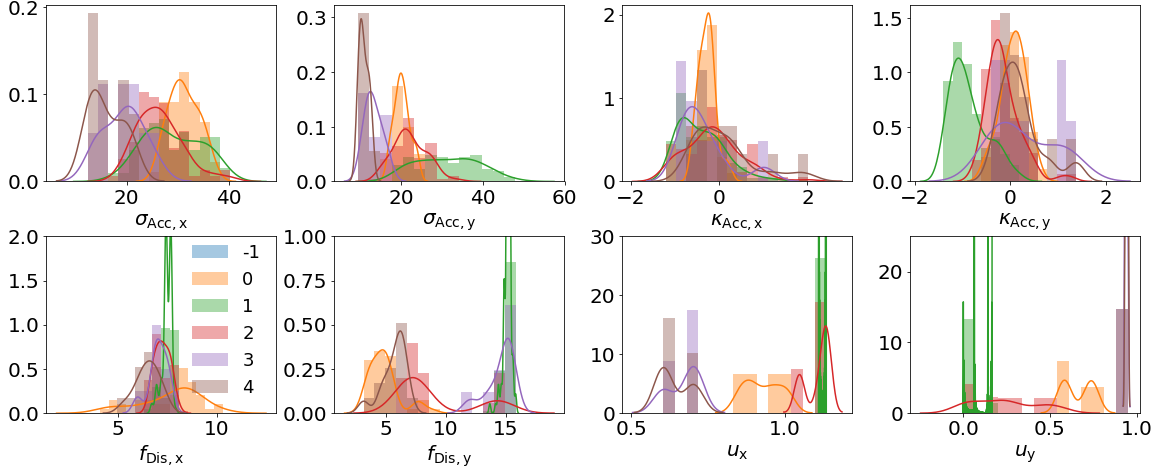}
\includegraphics[width=\textwidth]{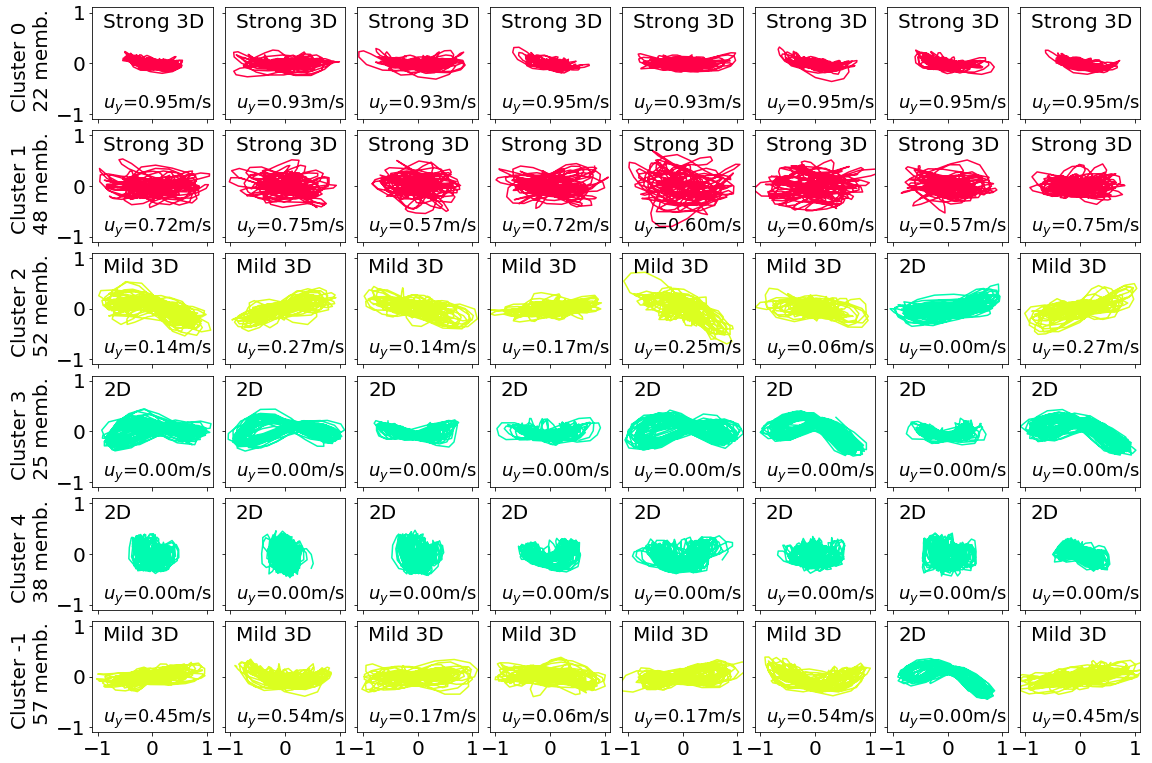}
\caption{{\it Upper panel:} Distributions of the statistical parameters of the clusters. The bars show the normalised histograms while the thin lines are the kernel density estimates. The cluster numbers correspond to the cluster members shown below. {\it Lower panel:} Examples of trajectories of cluster members. The colours and labels indicate the configuration. The last row shows examples of data points that are assigned to the noise category. The clusters are derived based on statistical properties of the accelerations and displacements, and not the trajectories shown here. The fact that the cluster members have visually similar trajectories indicate that the statistical parameters are descriptive of the physical scenarios.}
\label{fig:clusters}
\end{figure}

The upper panel of \figref{clusters} shows the parameter distributions for each of the clusters. The randomly selected cluster members shown in the lower panels have visually similar trajectories. Consequently, the statistical information is descriptive of the physical scenario, and can be used by the clustering algorithm do distinguish between the scenarios.


\subsubsection{Other clustering methods:}
In addition to the density based clustering, we tested other methods such as Gaussian Mixture and Agglomerative clustering. Gaussian mixture clustering is a generalisation of k-means clustering where it is assumed that all data points are generated from a mixture of a finite number of overlapping Gaussian probability distributions. 
Agglomerative clustering is a bottom-up hierarchical clustering approach, where each data point starts as its own cluster, and the clusters are subsequently merged. 
When using only the parameters that have a high significance in the density based clustering, the results of the three methods are very similar. 

\section{Response prediction} 
Given the clear correlation between oscillation pattern, configuration and statistical parameters, it should be possible to use supervised learning to predict the statistical properties of the response from a simple input. 

The end goal is to predict the long term fatigue for realistic current profiles. Since the fatigue is strongly dependent on the response, we simplify the problem and here we compare the response prediction from traditional methods, such as VIVANA-FD, with a data driven random forest learning approach. We randomly selected a test case in the strong 3D configuration with bottom flow speed of $U_\mathrm{tot} = 1.105$\,\ms{} and sampling interval 2500-3000 for comparison.

\subsection{Traditional method with VIVANA-FD}
The flow speed varies in magnitude and direction along the length of the riser model as shown in \figref{geometry} (normalised by total flow speed). The empirical frequency domain VIV prediction program VIVANA-FD was used for the case study using discrete response frequencies.
The equation of dynamic equilibrium is defined as 
\begin{equation}
M\ddot{r} + C\dot{r} + Kr = R \, ,
\end{equation}
where $R$ represents the external forces, $M$ incorporates the structural and hydrodynamic added mass, $C$ describes structural and hydrodynamic damping, and $K$ is the stiffness matrix. $M$, $C$ and $R$ are functions of the response vector $r$, which necessitates an iterative solution scheme. It is assumed that the excitation and response are harmonic at identical frequencies. This type of stationary response is assumed by most of the empirical VIV softwares \cite{Passano2014,Vandiver2007,Triantafyllou1999}. The main hydrodynamic coefficients are the added mass, excitation and damping coefficients, which are generalised from VIV model test with 2D flows.

The 3D flow is normally converted to 2D by using the total velocity. The displacement prediction with the 2D flow profile is presented in \figref{regression} (lower panel, green line). VIVANA-FD predicts a single response frequency ($f_\mathrm{osc}=6.87$\,Hz) to dominate the responses along the length of riser model. Consequently, the response will be harmonic due to the assumption of the load model. 

\subsection{Data driven approach} \label{sec:supervised}
Instead of using a physics-based model, the purpose of data driven modelling is to fit the data with a highly flexible model. 

\subsubsection{Random forest regression:}
Random forest regression is based on decision trees with a technique called bootstrap aggregation (also known as bagging) \cite{Breiman2001}. Rather than using individual decision trees, the data is randomly sampled (with replacement) and a decision tree is trained for each sample.

In order for each feature/variable to contribute equally to the fitting process, the input and output features must be scaled individually to the same range. We performed a min-max scaling to the range [0, 1], before randomly splitting the data into a training sample (80\%) and a test sample (20\%). The input parameters were \dist{}, \utot{}, \ux{}, \uy{}, and the output parameters were the remaining parameters in \tabref{params}. However, there was a clear connection between the parameters with an important role in the clustering, and the parameters that could be well determined with the random forest model, so we restricted the output to \sigposx{}, \sigposy{}, \fposx{}, \fposy{}, \sigpostot{}.

We used a grid search to optimise the hyper-parameters to a depth of 10 layers in the decision trees and 400 estimators. For a 5-fold cross validation, the best mean squared error on the normalised training data was 0.0062 and 0.0060 on the normalised test data. That the model performed better on the test data than the training data is a good sign that it was not over-fitting. For the individual parameters the mean square error was varying from 0.002 to 0.009. 

\subsection{Results and comparison}
\begin{figure}[bt]
\centering
\includegraphics[width=0.9\textwidth]{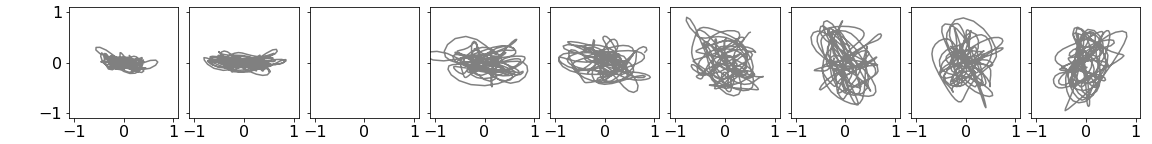}
\includegraphics[width=0.9\textwidth]{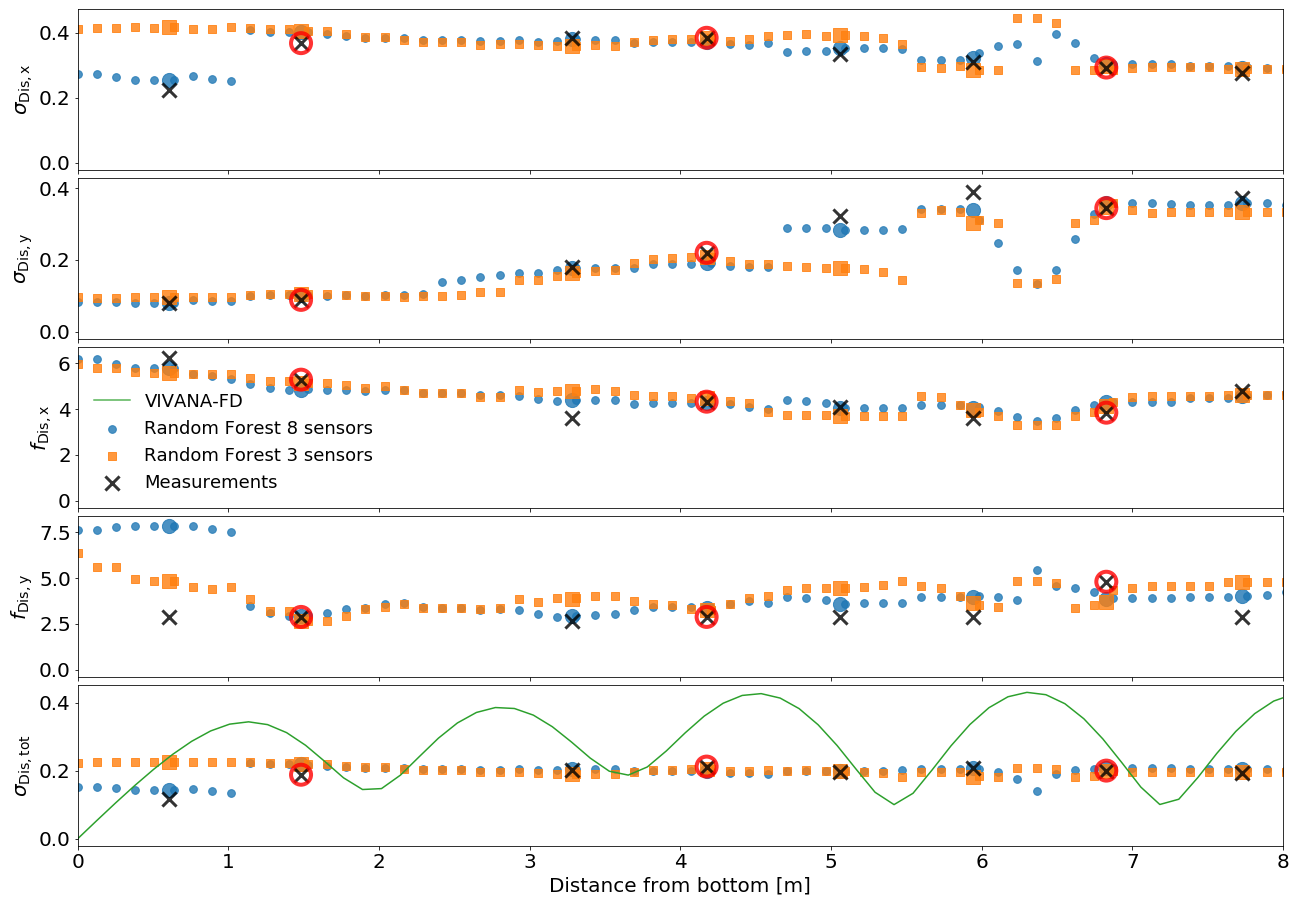}
\caption{{\it Top panel:} Trajectories of the case study; the strong 3D configuration with a bottom flow speed of 1.105\,\ms{} for sample range 2500-3000. {\it Subsequent panels:} The measurements (black crosses) and model predictions for the case study. The blue circles show the random forest prediction from fitting to data from all sensors with the sensor positions highlighted as larger circles. The orange squares show the predictions from fitting only three sensors (highlighted as red circles), but predicting for all sensors. The {\it lowest panel} shows the standard deviation of the displacement (in units of riser diameter). The green line is the prediction from VIVANA-FD assuming 2D flow.} 
\label{fig:regression}
\end{figure}

\figref{regression} shows the results of both predictive approaches compared to the measurements of the strong 3D configuration with bottom flow speed of $U_\mathrm{tot} = 1.105$\,\ms{} (excluded from training and test samples). As discussed in \secref{timeseries}, the measured variation of displacement may vary up to 15\% between the time intervals, and hence it is a reasonable estimate of the uncertainty on the measured displacement amplitudes plotted in the lowest panel of \figref{regression}.

Visually, the random forest model (blue) provide good predictions at the sensor locations. The sensor closest to the bottom is also closest to the pinned end. Consequently, it will have smaller displacements and the measurements are more prone to noise. Further analysis is required to identify the origin of the discrepancy and possible improvements of the model at this location. Around 6-7\,m from the bottom there is an "oscillation" in the model, mostly pronounced for \sigposy{}. Since there are no measurements between the accelerometers, it is uncertain whether this is an artefact of the model but the feature remains when fitting the model to fewer sensors.
 
Deep-water risers will normally have a limited number of sensors. Hence we tested the ability of the model to interpolate by fitting only to data from a sub-sample of the sensors (orange in \figref{regression}) but predicting the response at all sensor locations. The three-sensor fit leads to larger deviations than fitting all sensors, but still with decent performance, in particular for the predictions of the total displacement (lowest panel in \figref{regression}).

The lowest panel in \figref{regression} shows the variation of the total displacement. The VIVANA-FD prediction (green line) over-predicts the displacement amplitude relative to the measurements along the entire riser leading to a root mean squared error of 4.8 relative to the measurements at the sensor locations. In addition, the predicted response is harmonic compared to the almost flat behaviour of the measured responses. The explanation lies in the simplifying assumptions of VIVANA-FD. Firstly, the predicted responses are dominated by a single frequency leading to harmonic oscillations. This is contrary to the multi-frequency responses observed in the measurements (shown in \figref{timeseries_fft}). Secondly, the 2D assumption clearly leads to over-prediction of the displacement amplitude.


The random fores prediction is closer to the data both when fitted to all data (root mean squared error of 0.02 relative to sensor measurements) and to a reduced set of sensors. In the present case study, the VIVANA-FD model is significantly less accurate than the random forest model. This not surprising as it has been derived for a simplified scenario with 2D current but is compared to a 3D current scenario. However, the data driven method requires realistic training data, and without further information about the system, it cannot easily be transferred to e.g. a different type of riser. The ideal solution will be to combine the methods into a hybrid solution in order to obtain high precision and transferability.

\section{Summary and conclusion}
Using density based clustering we found a clear relation between dimensionality of the current and the observed pattern of riser movement and consequently fatigue. The response pattern can be identified from the statistical properties of the movement alone. We fitted a random forest model for the statistical parameters based on the local current conditions and position on the riser. The random forest model provide a more precise prediction of the displacement amplitude (\sigpostot{}) than the traditional approach using VIVANA-FD. However, as it is completely data driven, it does not provide any insights on the physics behind the riser response, and in order to translate the random forest model between different riser types and scale it to operational risers, additional training data is required spanning all relevant scenarios. The natural way forward is to combine the physics based modelling with the data driven approach in a hybrid solution.
%
%
%
\newline
\newline
{\bf Acknowledgements.} The authors would like to thank Anne Marthine Rustad for discussions and suggestions. This study was sponsored by Equinor, BP, Kongsberg Maritime, Trelleborg Offshore, Aker Solutions and Subsea7.
\bibliographystyle{splncs04}
\bibliography{bibliography}

\begin{thebibliography}{10}
\providecommand{\url}[1]{\texttt{#1}}
\providecommand{\urlprefix}{URL }
\providecommand{\doi}[1]{https://doi.org/#1}

\bibitem{Breiman2001}
Breiman, L.: Random forests. Machine Learning  \textbf{45}(1),  5--32 (Oct
  2001). \doi{10.1023/A:1010933404324}

\bibitem{Campello2013}
Campello, R.J.G.B., Moulavi, D., Sander, J.: Density-based clustering based on
  hierarchical density estimates. In: Pei, J., Tseng, V.S., Cao, L., Motoda,
  H., Xu, G. (eds.) Advances in Knowledge Discovery and Data Mining. pp.
  160--172. Springer Berlin Heidelberg, Berlin, Heidelberg (2013)

\bibitem{Hastie09}
Hastie, T., Tibshirani, R., Friedman, J.H.: The elements of statistical
  learning: data mining, inference, and prediction, 2nd Edition. Springer
  series in statistics, Springer (2009),
  \url{http://www.worldcat.org/oclc/300478243}

\bibitem{Lie2012}
Lie, H., Braaten, H., Jhingran, V., Sequeiros, O.E., Vandiver, K.:
  Comprehensive riser viv model tests in uniform and sheared flow. vol.~5. ASME
  2012 31st International Conference on Ocean, Offshore and Arctic Engineering
  (2012). \doi{10.1115/OMAE2012-84055}

\bibitem{Lie1998}
Lie, H., Mo, K., Vandiver, J.: Viv model test of a bare- and a staggered
  buoyancy riser in a rotating rig. Offshore Technology Conference,
  OTC-8700-MS, Houston, Texas (1998). \doi{10.4043/8700-MS}

\bibitem{McInnes2017}
McInnes, L., Healy, J., Astels, S.: hdbscan: Hierarchical density based
  clustering. The Journal of Open Source Software  \textbf{2}(11) (mar 2017).
  \doi{10.21105/joss.00205}

\bibitem{Passano2014}
Passano, E., Larsen, C., Lie, H., , Wu, J.: VIVANA - Theory Manual Version 4.4
  (2014)

\bibitem{Srivilairit2009}
Srivilairit, T., Manuel, L.: Vortex-induced vibration and coincident current
  velocity profiles for a deepwater drilling riser. Journal of Offshore
  Mechanics and Arctic Engineering  \textbf{131} (2009).
  \doi{10.1115/1.3058684}

\bibitem{Thorsen2014}
Thorsen, M., S\ae{}vik, S., Larsen, C.: A simplified method for time domain
  simulation of cross-flow vortex-induced vibrations. Journal of Fluids and
  Structures  \textbf{49},  135 -- 148 (2014).
  \doi{10.1016/j.jfluidstructs.2014.04.006}

\bibitem{Thorsen2017}
Thorsen, M., S\ae{}vik, S., Larsen, C.: Non-linear time domain analysis of
  cross-flow vortex-induced vibrations. Marine Structures  \textbf{51},  134 --
  151 (2017). \doi{10.1016/j.marstruc.2016.10.007}

\bibitem{Triantafyllou1999}
Triantafyllou, M., Triantafyllou, G., Tein, Y.D., Ambrose, B.D.: Pragmatic
  riser viv analysis. Offshore Technology Conference, OTC-10931-MS, Houston,
  Texas (1999). \doi{10.4043/10931-MS}

\bibitem{TRIM2005335}
Trim, A., Braaten, H., Lie, H., Tognarelli, M.: Experimental investigation of
  vortex-induced vibration of long marine risers. Journal of Fluids and
  Structures  \textbf{21}(3),  335 -- 361 (2005).
  \doi{10.1016/j.jfluidstructs.2005.07.014}, marine and Aeronautical
  Fluid-Structure Interactions

\bibitem{Vandiver2000}
Vandiver, J.: Predicting lock-in on drilling risers in sheared flows.
  Flow-Induced Vibration Conference, Lucerne, Switzerland (2000)

\bibitem{Vandiver2007}
Vandiver, J., Li, L.: Shear7 v4.5 Program Theoretical Manual (2007)

\bibitem{Vandiver2006}
Vandiver, J.K., Swithenbank, S.B., Jaiswal, V., Jhingran, V.: Fatigue damage
  from high mode number vortex-induced vibration. vol.~4. ASME 2006 25th
  International Conference on Ocean, Offshore and Arctic Engineering (2006).
  \doi{10.1115/OMAE2006-92409}

\bibitem{Westfall2014}
Westfall, P.H.: Kurtosis as peakedness, 1905 - 2014. r.i.p. American
  Statistician  \textbf{68}(3),  191--195 (aug 2014).
  \doi{10.1080/00031305.2014.917055}

\end{thebibliography}
\end{document}